\documentclass[12pt]{iopart}
\usepackage{iopams} 
\usepackage{setstack}
\usepackage{graphicx}
\usepackage{xcolor}
\begin{document}

\title[Matrix product solution for a partially asymmetric 1D lattice gas with a free defect]{Matrix product solution for a partially asymmetric 1D lattice gas with a free defect}

\author{Ivan Lobaskin$^1$, Martin R Evans$^1$ and Kirone Mallick$^2$}

\address{$^1$ School of Physics and Astronomy, University of Edinburgh, Peter Guthrie Tait Road, Edinburgh, EH9 3FD, United Kingdom}
\address{$^2$ Institut de Physique Th{\' e}orique, Universit{\' e} Paris-Saclay, CEA and CNRS, 91191 Gif-sur-Yvette, France}
\eads{\mailto{ivan.lobaskin@ed.ac.uk}, \mailto{m.evans@ed.ac.uk}, \mailto{kirone.mallick@ipht.fr}}
\vspace{10pt}

\begin{abstract}
A one-dimensional, driven lattice gas with a freely moving, driven defect particle is studied.
Although the dynamics of the defect are simply biased diffusion, it disrupts the local density of the gas, creating nontrivial nonequilibrium steady states.
The phase diagram is derived using mean field theory and comprises three phases.
In two phases, the defect causes small localized perturbations in the density profile.
In the third, it creates a shock, with two regions at different bulk densities.
When the hopping rates satisfy a particular condition (that the products of the rates of the gas and defect are equal), it is found that the steady state can be solved exactly using a two-dimensional matrix product ansatz. 
This is used to derive the phase diagram for that case exactly and obtain exact asymptotic and finite size expressions for the density profiles and currents in all phases.
In particular, the front width in the shock phase {on a system of size $L$ is found to scale as $L^{1/2}$}, which is not predicted by mean field theory.
The results are found to agree well with Monte Carlo simulations.
\end{abstract}

\section{Introduction}
Minimal models of one-dimensional, driven diffusive gases have been used to model a wide variety of systems, including traffic flow \cite{wolf1996traffic}, diffusion in narrow channels \cite{cividini2017driven}, mRNA translation \cite{SNCM2018,scott2019power} and as discrete versions of fluid equations, such as Burgers' equation \cite{derrida1993exact}.
Despite their simplicity, their study has uncovered a rich phenomenology, with phase transitions and non-trivial many body phenomena such as shock formations \cite{janowsky1992finite,derrida1993exact,derrida1997shock,mallick1996shocks,jafarpour2005exact,tabatabaei2006shocks} and condensation \cite{evans1996}.
At the same time, the techniques used to study them have enriched the understanding of nonequilibrium steady states (NESSs), by providing a multitude of examples in which the NESS can be solved exactly \cite{blythe2007nonequilibrium,CMZ2011}.

One of the most studied classes of such systems are simple exclusion processes (SEPs).
These are models of gases in which particles hop stochastically on a lattice and interact locally via hard-core repulsion.
They are usually divided into totally asymmetric (TASEP), partially asymmetric (PASEP) and symmetric (SSEP) cases.
Many exact results have been derived for the steady states of these systems.
For  comprehensive reviews, we direct the reader to \cite{blythe2007nonequilibrium,CMZ2011}.

As a generalization, one may introduce a defect particle that hops at different rates to the ``normal" (environment) particles and which may overtake or be overtaken by them, see e.g. \cite{burlatsky1992directed,derrida1993exact,evans1996,mallick1996shocks,miron2020phase}.
Physically, one can think of the defect as being driven by an external field that does not affect the other particles, or it may be affected by a local force, such as from optical tweezers.
Particle overtaking may be facilitated for example by a finite but narrow channel width \cite{miron2020phase}.
Then particles may overtake each other but collisions during this process will alter their drift speed.
At the level of the minimal model, exchange is simply modelled as a stochastic process with some effective rates.

The most general formulation of this problem has six independent rates: left and right hopping rates for the environment particles and the defect, and left and right overtaking rates.
This general case currently remains beyond the reach of exact calculation methods.

In this paper we address a novel solvable limit, that of a PASEP with a generalized first class defect.
By ``first class", we mean that the defect has a higher priority than the  environment (second class) particles in the dynamics.
It may overtake environment particles with its usual hopping rates, but it cannot be overtaken by them.
It is ``generalized" in the sense that its hopping rates are different to that of the environment.
It might appear at first glance that this model should be very simple, given the trivial dynamics of the defect. 
However, it turns out to be nontrivial both in its physical behaviour and the mathematical structure of its exact solution.

We now briefly comment on related models that have been studied in the literature.
The first exact solution of a TASEP was the open boundary case \cite{derrida1992exact}.
Thereafter it was shown that this solution can be expressed as a matrix product \cite{dehp1993}, which has since become one of the main approaches for obtaining exact results.
Although the open boundary case is not directly related to the defect problem on a ring, it is interesting to compare, as it has some similar phenomenology, like shock formations \cite{schutz1993phase,tabatabaei2006shocks}.
Some exact results for shocks have also been obtained in ASEPs with a site defect \cite{janowsky1992finite,janowsky1994exact,basu2016passage}.

The earliest instance of an ASEP with defect that was solved exactly was that of a second class defect  \cite{derrida1993exact,derrida1997shock}.
A second class particle does not affect the dynamics of the environment particles but its own dynamics depend on the local density.
This property means that it can be used to track the location of shock fronts.
This system was solved exactly in the TASEP \cite{derrida1993exact} and PASEP \cite{derrida1997shock}.
An exact solution was also obtained for a TASEP in which the defect has different hopping and overtaking rates to the environment \cite{mallick1996shocks}.
In this case, the defect does affect the environment and the phase diagram depends on the defect parameters.

Another limit that has received attention is that of the symmetric environment with a totally or partially asymmetric defect.
The case without overtaking has a long history in literature, where it was mainly studied using mean field methods \cite{burlatsky1992directed,burlatsky1996motion}.
Recently, the case without overtaking was solved exactly \cite{lobaskin2020driven,ayyer2020simple} and the case with overtaking was analyzed using mean field theory \cite{cividini2018driven, miron2020phase}.

A related class of systems that has received some attention in the literature is that of a dynamic blockage \cite{sahoo2014dynamic,sahoo2016asymmetric,szavits2020current}.
Here, overtaking is typically not considered but the defect particle (blockage) is allowed to diffuse on the lattice, as well as perform non-local jumps, modelling a binding-unbinding process.
Dynamic blockages are of particular interest for the modelling of DNA transcription, where proteins that spontaneously bind and unbind from the DNA, can act as blockages for the transcribing RNA-polymerase.
This has been studied numerically for a TASEP on a ring \cite{sahoo2014dynamic} and with open boundaries \cite{sahoo2016asymmetric} and some exact results have been obtained in the low and high density limits \cite{szavits2020current}.

Perhaps the closest to the system considered in this paper that has been solved exactly is a partially asymmetric environment with a totally asymmetric defect and overtaking \cite{sasamoto2000one}.
Though we remark that the total asymmetry of the defect changes the exact solution considerably.

The remainder of this paper is structured as follows.
In \sref{mft}, we define the model and derive the phase diagram using mean field analysis.
From \sref{model-matrix} onward, we restrict ourselves to a special case, in which it is found that the steady state can be solved exactly using a matrix product ansatz.
As the matrix product ansatz presented here has some particular features, we include a proof that it gives the correct steady state measure in \ref{proof}.
In \sref{phase-diagram}, we use the matrix product formulation to re-derive the phase diagram.
In \sref{asymptotics}, we derive asymptotic expressions for the density profiles and currents using a saddle point approximation.
This shows that the picture presented through the mean field analysis is qualitatively correct.
In \sref{exact}, we state the exact finite size expressions for the nonequilibrium partition function, density profiles and currents and examine the symmetric and totally asymmetric limits.
In \sref{conclusion}, we make some concluding remarks and comment on the remaining open questions related to this problem.

\section{Mean field theory} \label{mft}
\subsection{Model definition}
We consider an $L+1$ site ring with $M+1$ particles hopping stochastically in continuous time. 
$M$ of these particles (the environment) hop to the right with rate $p$ and to the left with rate $q$.
The remaining particle (the defect) hops to the right with rate $p'$ and to the left with rate $q'$.
It is convenient to define the {environment} asymmetry parameter 
\begin{equation}
x = \frac{q}{p}\;,
\end{equation}
the drift speeds of the environment and defect {respectively as} 
\begin{equation}
   v=p-q \; ,\qquad v'=p'-q', 
\end{equation}
and the mean environment density 
\begin{equation}
  \rho =\frac{M}{L}\;.  \label{rho}
\end{equation}

The particles interact via simple exclusion.
Environment particles may not overtake other environment particles or the defect.
The defect particle may overtake environment particles (to the left or to the right) with its usual hopping rates.
Thus the defect particle does not distinguish between environment particles and empty sites.
Its dynamics are that of a free particle with left and right hopping rates $p',\; q'$.
However, the environment particles cannot jump over the defect, which means that the defect can disrupt the environment and has a nontrivial effect on its density profile.
The dynamics of the system can be represented schematically as follows:
\begin{eqnarray*}
    10
    \underset{q'}{\overset{p'}{\rightleftarrows}}
    01 \quad ; \quad
    12
    \underset{q'}{\overset{p'}{\rightleftarrows}}
    21 \quad ; \quad
    20
    \underset{q}{\overset{p}{\rightleftarrows}}
    02 \quad ;
\end{eqnarray*}
\newline
where 0, 1, 2 represent empty sites, the defect particle and environment particles respectively.
Note that unlike some recent papers, the defect is denoted by 1 and the environment particles by 2, because in this case the defect is a ``first class" particle.

The system has the symmetry $p\leftrightarrow q,\; 0\leftrightarrow 2$ (equivalently $x\to 1/x,\; v\to -v,\; \rho \to 1-\rho$).
With this in mind, we will, without loss of generality, set $p>q$, unless otherwise stated.
This means that by default, the net current of environment particles is taken to be to the right.

\subsection{Mean field analysis}
The phase diagram for this model can be obtained from mean field theory.
We consider the system in the reference frame of the defect, so that its position may always be taken to be $L+1$.
Let $n_k$ be the probability that the $k$-th site to the right of the defect is occupied by an environment particle.
Then at the mean field level (i.e. approximating two point correlation functions as a product of one point functions), the probabilities evolve under the following equations in the bulk ($1<k<L$) and near the defect respectively:
\numparts
\begin{eqnarray}
    \fl
    \frac{\partial n_k}{\partial t} = -p n_k (1-n_{k+1}) -q n_k (1-n_{k-1}) - (p' + q')n_k \nonumber\\
    + p n_{k-1}(1-n_k) + q n_{k+1} (1-n_k) +q' n_{k-1} + p' n_{k+1}\\
    \fl
    \frac{\partial n_1}{\partial t} = -p n_1 (1-n_{2}) - (p' + q')n_1 + q n_{2} (1-n_1) 
    + q' n_{L} + p' n_{2} \\
    \fl
    \frac{\partial n_L}{\partial t} = - q n_L (1-n_{L-1}) - (p' + q')n_L
    + p n_{L-1}(1-n_L) + q' n_{L-1} + p' n_{1} \; .
\end{eqnarray}
\endnumparts
The steady state is characterized by a constant current of environment particles:
\numparts
\begin{eqnarray}
    J' = p n_k (1-n_{k+1}) - q n_{k+1} (1-n_k) - p' n_{k+1} + q' n_k \; , \; 1\leq k <L \\
    J' = - p' n_{1} + q' n_L \label{current}
\end{eqnarray}
\endnumparts
where $J'(={\rm const.})$ is the steady state current of environment particles in the reference frame of the defect.
The current in the stationary frame will be denoted by $J$. 
We remark that \eref{current} is in fact an exact expression as it only involves one point functions and therefore does not require the mean field assumption.
Since the defect moves as a free particle, $J$ is related to $J'$ simply by a Galilean boost:
\begin{equation}
    J=J'+\rho v' \;. \label{galileo}
\end{equation}
This is because the defect moves with average speed $v'$, independently of its local environment.
In the stationary frame, the defect is equally likely to be anywhere due to translational invariance.
Therefore, the average effect of the boost on the current will be $\rho v'$, where $\rho$ is the (global) average density.

Letting $y=k/L$ ($0< y \leq 1$), in the continuum limit ($L\to \infty$) we get the following boundary value problem for the density profile in the steady state:
\numparts
\begin{eqnarray}
    D \frac{\partial n(y)}{\partial y} = 
    vn(y)(1-n(y)) - v'n(y) -J' \label{DE}\\
    J' = -p' n(0) + q' n(1) \; , \label{BC}
\end{eqnarray}
\endnumparts
where $D = \frac{1}{2L}(p+q+p'+q')$. 
The solutions to  (\ref{DE}) are $\coth$ and $\tanh$ functions with characteristic lengths scaling as $\xi \propto D \sim L^{-1}$.
(From exact results for a specific choice of parameters, this scaling will be shown to be correct when the effects of the defect are localized, but when the defect causes a shock, the characteristic length scales as $\xi \sim L^{-1/2}$ instead.)
Then $J'$ and a constant of integration have to be fixed using the boundary condition \eref{BC} and the global particle number constraint.
This calculation becomes quite cumbersome and not very informative so instead we give a qualitative picture by considering the fixed points (i.e. uniform solutions) of \eref{DE}.

\subsection{Phase diagram}
The fixed points of (\ref{DE}) satisfy a quadratic equation, 
\begin{equation}
    vn(1-n) - v'n -J'=0 \label{quadratic} 
\end{equation}    
which can be readily solved to give two plateau  densities
\begin{equation}
n_{\pm} (J')= \frac{v-v' \pm \sqrt{(v-v')^2 -4vJ'}}{2v} \; .\label{npm}
\end{equation}
In particular,
\begin{equation}
n_+ + n_- = 1 - \frac{v'}{v}\;. 
\end{equation}

\begin{figure}
    \centering
    \includegraphics[scale=0.6]{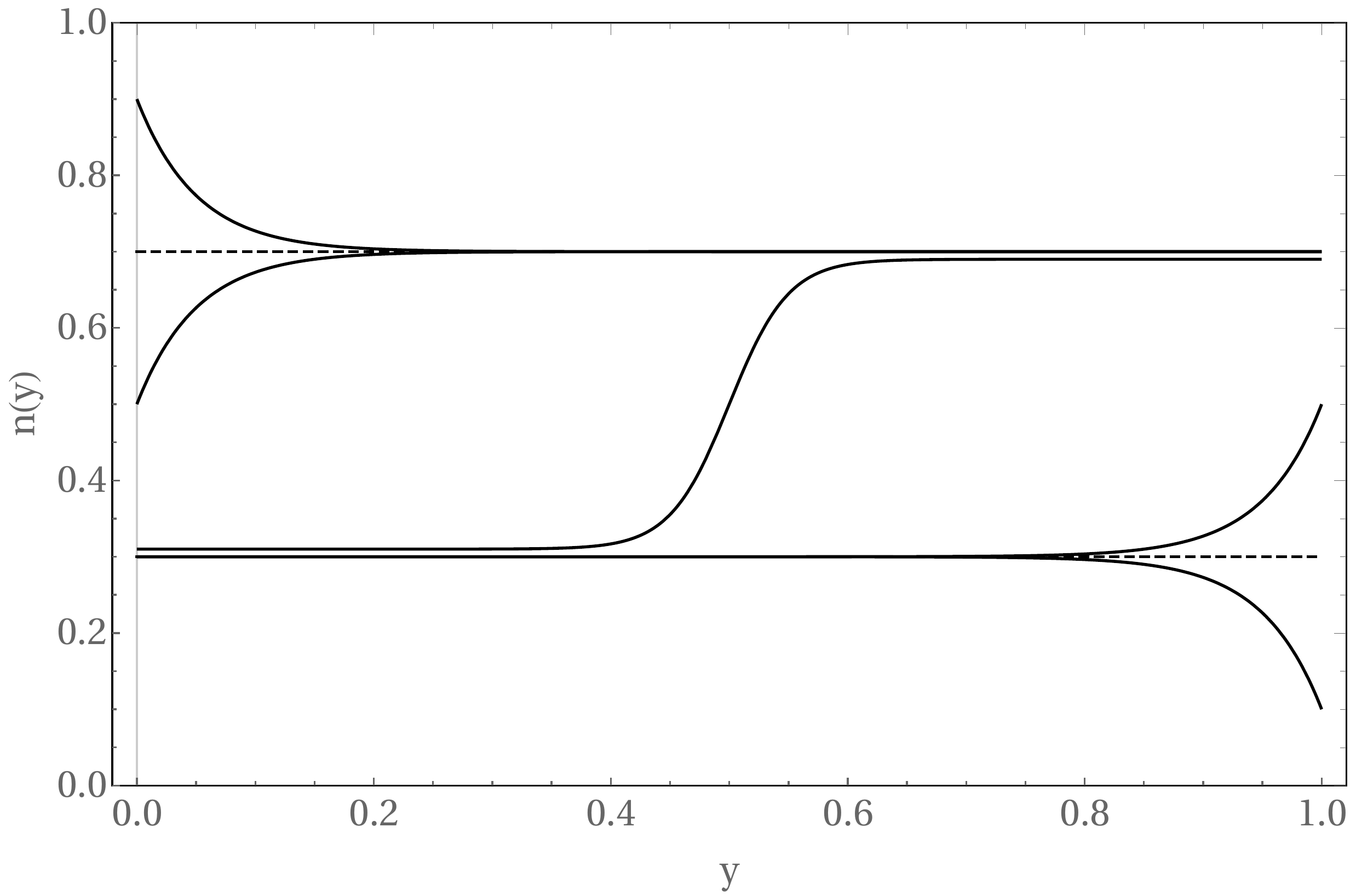}
    \caption{Schematic representation of mean field environment density profiles in the reference frame of the defect.
    Three types of solutions are observed: localized decays to the left or right of the defect and a shock profile with two bulk densities. 
     The solutions shown are for $v>0$ (net environment current to the right).
    In the opposite case, the orientations of the profiles are flipped.}
    \label{fig:mft_flows}
\end{figure}

The RHS of \eref{DE} will be positive for $n_-(J')<n(y)<n_+(J')$ and negative for $n(y)<n_-(J')$ and $n(y)>n_+(J')$.
By considering how a profile with some boundary values $n(0),n(1)$ will behave, we see that there exist three types of solutions.
In the first case, the density profile starts at some boundary value $n(0)>n_-(J')$ and decays exponentially over a length $O(1/L)$ to bulk density $n_+(J')\approx \rho$, where $\rho$ is given by (\ref{rho}). Thus in the limit $L \to \infty$,  $n(1) = n_+(J')=  \rho$.
The second case is similar to the first but with a bulk density of $n_-(J') \approx  \rho$ and the exponential profile occurring at the other boundary (with $n(1)<n_+(J')$).
Thus in the limit $L \to \infty$,  $n(0) = n_-(J')=  \rho$.
In the third case, the density profile starts very close to $n_-(J')$ and moves to $n_+(J')$ at some intermediate value of $y$.
Recalling the earlier comments on the forms and scaling of the solutions, we see that this will be a $\tanh $ profile that becomes a sharp shock in the limit $L\to \infty $.
The three types of solutions are shown schematically in \fref{fig:mft_flows}.

Thus the defect creates either small localized perturbations to its right ($\rho \approx n_+(J')$) or left ($\rho \approx n_-(J')$), or a shock ($n_-(J')<\rho <n_+(J')$) in the environment density profile.
We will correspondingly call these the right/left localized phases and the shock phase.

\subsubsection{Right localized phase}

{In the right localized solution we have 
\begin{equation}
 n_+ =\rho \; , \qquad J' = v\rho(1-\rho) -v' \rho \; , \qquad  n_- =1-\rho- \frac{v'}{v} \; . \label{rl}
\end{equation}
In order for the right localized solution to be consistent we must have
\begin{equation}
n(0)>n_-\;.\label{rlcon}
\end{equation}
Equation (\ref{BC}) implies 
\begin{equation}
   n(0) =\rho\frac{q'}{p'} -  \frac{J'}{p'} = \rho -\rho(1-\rho)\frac{v}{p'}\;. \label{rln0}
\end{equation}
From  \eref{rlcon}, \eref{rl}, \eref{rln0} we then obtain the condition for the right localized phase
\begin{equation}
    \rho > \frac{v - 2p' + \sqrt{v^2 + 4p'q'}}{2v} \; . \label{rho1}
\end{equation}}

\subsubsection{Left localized phase}
{In the left localized solution we have 
\begin{equation}
 n_- =\rho \; , \qquad J' = v\rho(1-\rho) -v' \rho \; , \qquad
n_+ =1-\rho- \frac{v'}{v} \; .\label{rlnm}
\end{equation}
In order for the left localized solution to be consistent we must have
\begin{equation}
n(1)< n_+ \; .
\end{equation}
Equation (\ref{BC}) implies 
\begin{equation}
n(1) = \rho +\rho(1-\rho)\frac{v}{q'}. 
\end{equation} 
We then obtain the condition for the left localized phase
\begin{equation}
    \rho <  \frac{v + 2q' -\sqrt{v^2 + 4p'q'}}{2v} \; .\label{rho2}
\end{equation}}

\subsubsection{Shock phase}
In the shock phase  we simultaneously have $n(0)=n_-(J')$ and $n(1)=n_+(J')$.
Thus the current satisfies:
\begin{equation}
    J' = -p' n_-(J')+q' n_+(J') \; . \label{jshock}
\end{equation}
As before, we can eliminate $J'$ to get expressions for the  densities  to the left and right of the shock
\begin{equation}
    \rho _{\pm} = \frac{v \mp 2p'_{\pm} \pm \sqrt{v^2 + 4p'q'}}{2v} \; , \label{critical_density}
\end{equation}
where $p'_+=p'$ and $p'_-=q'$.
From \eref{jshock}, this gives a current
\begin{equation}
    J' = - \frac{v'}{2} -\frac{1}{2v}(4p'q'+(p'+q')\sqrt{v^2+4p'q'}) \; . \label{jshock2}
\end{equation}
The density profile is
\begin{eqnarray}
    n(y)=
    \left\{
    \begin{array}{ll}
        \rho_-, \quad y<y_0 \\
        \rho_+, \quad y>y_0 
    \end{array}
    \right.
    \; ,
\end{eqnarray}
where $y_0$ is the location of the shock, which must satisfy
\begin{equation}
    \rho _- y_0 + \rho _+ (1-y_0) = \rho \; .
\end{equation}
This can evidently only hold when $\rho_- <\rho <\rho_+$, which implies that $\rho _\pm$, as given by \eref{critical_density} are the densities at which the transitions from the localized to the shock phase occur.
The densities (\ref{critical_density}) coincide with the critical densities  \eref{rho1},\eref{rho2} for the existence of the right and left localized  phases (as expected).

To understand the phase diagram, we consider  fixed $p,q,q'$ and vary $p'$ and $\rho$.
{On increasing the density $\rho$ we have a transition at $\rho=\rho_-$ from the 
left localized phase to the shock phase and a transition when $\rho= \rho_+$ from the shock phase to the
right localized phase. As $\rho$ increases through the shock phase from
$\rho_-$ to $\rho_+$, the position of the shock moves from  $y=1$ to $y=0$. With regard to $p'$, the left and right localized phases occur in the small and large $p'$ limits respectively and are separated by the shock phase.}

\begin{figure}
    \centering
    (a)\includegraphics[scale=0.55]{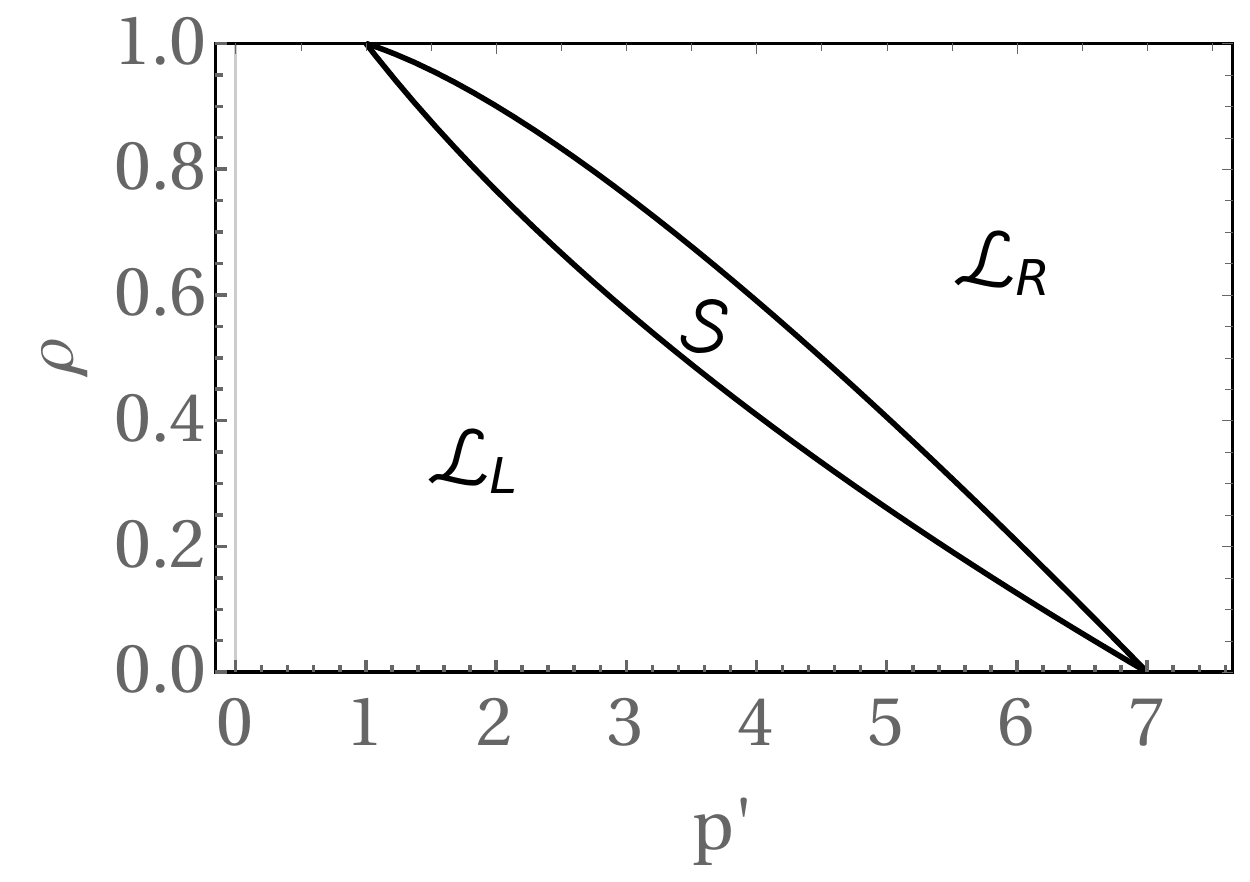}
    (b)\includegraphics[scale=0.55]{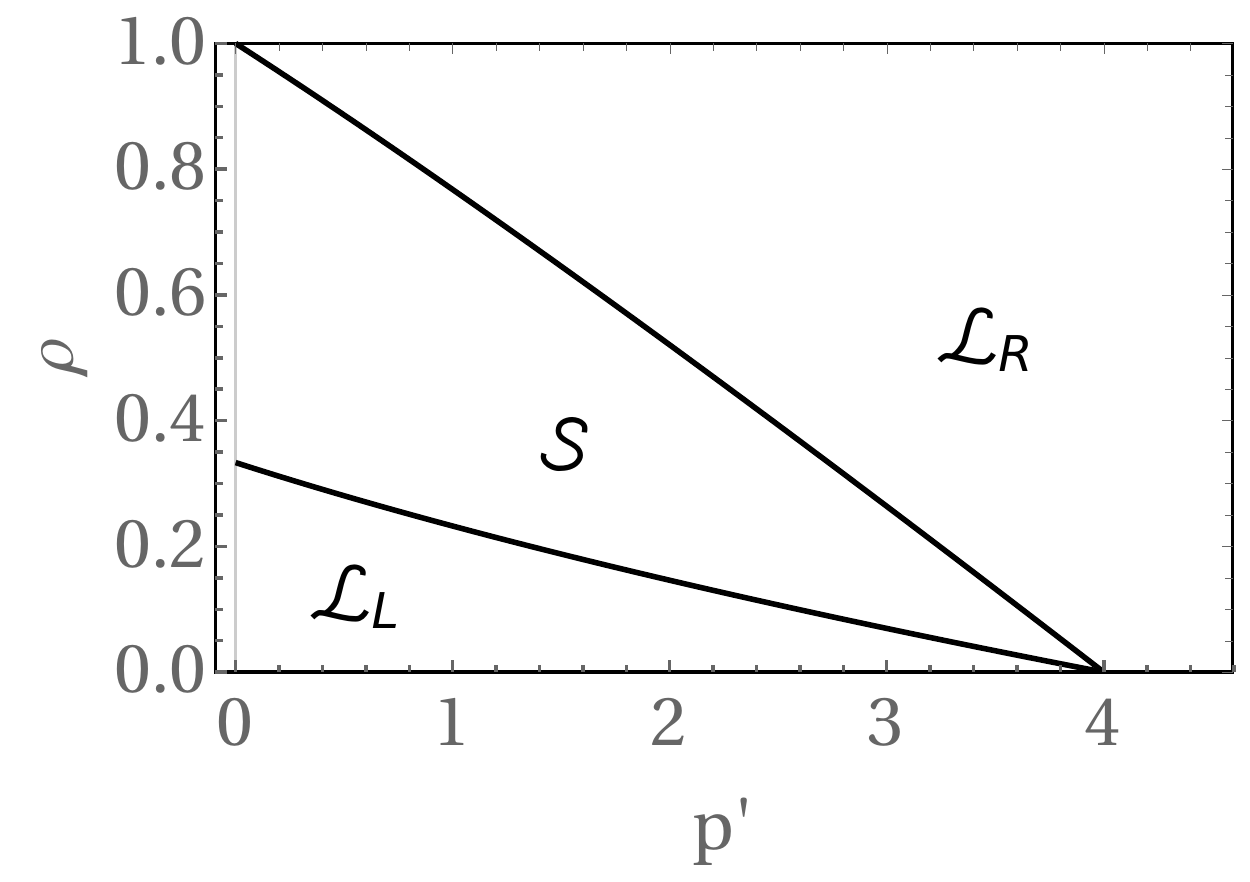} 
    \caption{Phase diagrams with the defect right hopping rate $p'$ and environment density $\rho$ as control parameters.
    The left and right localized phases (${\cal L}_{L/R}$) are separated by a shock phase (${\cal S}$) in each case (the shock phase vanishes for $v=0$, which is not shown here).
    The left localized phase shrinks as $q'$ is decreased and vanishes in the totally asymmetric defect limit ($q'\to 0$).
    The fixed parameters are: (a) $p=4,q=1,q'=4$; (b) $p=4,q=1,q'=1$.
    }
    \label{fig:mft_phase}
\end{figure}

{The phase boundaries are  delimited by the curves given by \eref{critical_density} (see \fref{fig:mft_phase}).
These two curves intersect at $\rho=0,\; v'=v$ and $\rho=1,\; v'=-v$ (switched if $v<0$).
Thus there is always a finite region in which the shock phase exists, except when $v=0$ (the symmetric environment limit).}
 The fact that shocks exist only when $-v<v'<v$ can also be stated as a consistency condition for the critical densities (which follows from \eref{critical_density}):
\begin{equation}
    \rho _+>\rho _- \Leftrightarrow |v'|<|v| \; .
\end{equation}

This helps to qualitatively understand the phase diagram. 
The inequality $v'<v$ means that the defect has to be slower (has a smaller drift) than the environment.
This is reminiscent to the mechanism that causes shocks in a TASEP with a slow defect \cite{mallick1996shocks}.
At the same time, we note that a shock profile for environment particles corresponds to a flipped shock profile for holes.
Holes can be considered to have a drift velocity $-v$, so for a shock to exist, we must also require $v'>-v$, giving the other inequality.

\subsection{Current  in stationary frame}
From the above analysis, in particular \eref{galileo}, \eref{rl}, \eref{rlnm}, \eref{jshock2}, we obtain mean field expressions for the current of environment particles in the stationary frame:
\begin{eqnarray}
    \fl J =
    \left\{
    \begin{array}{ll}
        \rho (1-\rho )v, & {\rm localized \; phases} \\
        (\rho - 1/2)v' -\frac{1}{2v}(4p'q'+(p'+q')\sqrt{v^2+4p'q'}), \quad & {\rm shock \; phase}
    \end{array}
    \right.
    \; . \label{mft_current}
\end{eqnarray}
In the localized phases, the current is the same as for a pure (defectless) PASEP.
This is to be expected since the effects of the defect are only felt locally.
In the shock phase, the defect has a throttling effect on the environment current.
This is especially apparent if one considers $J'=J-\rho v'$, which becomes independent of density in the shock phase.
The effect of the defect can be understood as preventing the environment current from rising above a certain threshold (see \fref{fig:current}).

These mean field results will be shown to agree with the asymptotic results obtained from an exact solution for a specific choice of parameters in \sref{asymptotics}.

\section{Exact solution for steady state using matrix product formulation} \label{model-matrix}
The steady state can be solved exactly through a matrix product ansatz if the hopping rates satisfy the condition 
\begin{equation}
  pq=p'q'\;.  
\end{equation}
We will restrict ourselves to this case for the remainder of this paper.
We can then write the hopping rates of the defect in terms of a single defect parameter, $\alpha $, by setting $p'=\alpha p$ and $q'= q/\alpha $.

The full state of the system is specified by a string $\{ \tau _1, \tau _2,\dots ,\tau _{L+1}\}$, where the variables $\tau _i$ can take values $0,1,2$ depending on whether the site is unoccupied, occupied by the defect or occupied by an environment particle respectively.
At long times, the system reaches a nonequilibrium steady state.
We denote the unnormalized weight of a configuration in the steady state by $f(\{\tau _1,\dots ,\tau _{L+1} \})$.
Then the probability of that configuration is
\begin{equation}
    P(\{\tau _1,\dots ,\tau _{L+1} \}) = 
    \frac{f(\{\tau _1,\dots ,\tau _{L+1} \})}{Z_{L,M}} \; ,
\end{equation}
where the normalization (also called a nonequilibrium partition function) is defined as
\begin{equation}
    Z_{L,M} = \sum\limits _{\{\tau \}} f(\{\tau _1,\dots ,\tau _{L+1} \}) \; ,
\end{equation}
in which the sum is over all configurations.

The steady state of this system is of the matrix product type (see \cite{blythe2007nonequilibrium,wood2020combinatorial} for reviews of the models that have been solved and the combinatorial mappings of the solutions).
In other words, there exist matrices $X_0,X_1,X_2$, such that the weight of any configuration is given by the corresponding matrix product:
\begin{equation}
    f(\{\tau _1,\dots ,\tau _{L+1} \}) = \tr (X_{\tau _1}\dots X_{\tau _{L+1}} ) \; .
\end{equation}
The use of the trace reflects the translational symmetry of the steady state.
For the problem at hand, the following matrices are found to work:
\begin{eqnarray}
    X_0 = 
    \left( \begin{array}{cc}
        \alpha & 0 \\
        \alpha & 1
    \end{array} \right) , \qquad
    X_1 =
    \left( \begin{array}{cc}
        0 & 1 \\
        0 & 1
    \end{array} \right) , \qquad 
    X_2 = 
    \left( \begin{array}{cc}
        \alpha  & 0  \\
        \alpha  & x
    \end{array} \right)
    . \label{matrices}
\end{eqnarray}

The proof that the weights generated by them give the correct steady state is given in \ref{proof}.
We now make a few remarks on this solution in the context of other known matrix product states.

First, it is somewhat unusual these these matrices are finite dimensional, as often infinite dimensional matrices are required, for instance for the open boundary TASEP \cite{dehp1993}.
Two-dimensional representations have been found for some ASEPs, including a fine-tuned case of a shock in a PASEP \cite{derrida1997shock} and open systems in parallel updating schemes \cite{jafarpour2005exact}.

Secondly, the matrices do not form a closed algebra.
Instead, we have:
\numparts
\begin{eqnarray}
    \alpha p X_1 X_0 - q/\alpha X_0 X_1 = x_0 X_1 - Y_0 \label{algebra10} \; , \\
    \alpha p X_1 X_2 - q/\alpha X_2 X_1 = x_2 X_1 - Y_2 \label{algebra12} \; , \\
    p X_2 X_0 - q X_0 X_2 = x_0 X_2 - x_2 X_0 \; , \label{algebra20}
\end{eqnarray}
\endnumparts
where:
\begin{eqnarray}
    \fl
    x_0 = p(\alpha - x) , \;
    x_2 = px(\alpha -1) , \;
    Y_0 = p
    \left( \begin{array}{cc}
    -\alpha ^2  & 0 \\
    -\alpha ^2  & x/\alpha 
    \end{array} \right) , \;
    Y_2 = p\left( \begin{array}{cc}
     -\alpha ^2  & 0 \\
    -\alpha ^2  & x^2/\alpha 
    \end{array} \right) .
\end{eqnarray}
We note that out of these relations, only (\ref{algebra20}) is expressed solely in terms of the original matrices $X_0,X_1,X_2$.
This means that the usual proof that the matrices form a steady state solution \cite{blythe2007nonequilibrium} does not follow automatically; we give a modified  proof in \ref{proof}.
We give a more robust argument that no representation exists for this problem that allows reductions similar to \eref{algebra20} for all pairs of matrices in \ref{reductions}.

Finally, it can be checked that this solution only works for systems with exactly one defect.
Exact diagonalizations of small systems with two defects suggest that if a matrix product ansatz exists for that case, it is significantly more complicated than the one presented here.
Moreover, preliminary results from an investigation of the Yang-Baxter integrability of this system suggest that the single defect case is integrable whereas the many defect case is not \cite{tobepublished}.

\section{Exact phase diagram}\label{phase-diagram}
\subsection{Nonequilibrium partition function}
We now use the matrix product formulation to calculate the nonequilibrium partition function and hence the exact phase diagram.
It is helpful to consider the matrix
\begin{equation}
    C = X_0 + z X_2 \; , \label{C}
\end{equation}
where $z$ can be thought of as the fugacity of environment particles.
This gives an easy way to enumerate all configurations of the system.
First, we fix the location of the defect, which can be done without loss of generality due to the translational symmetry of the steady state.
Then we use the matrix $C$ for the $L$ remaining sites, which allows for the possibility of each site being empty or occupied by an environment particle.
Finally, we take the trace and pick out the terms with exactly $M$  environment particles, which corresponds to isolating the coefficient of $z^{M}$.
This can be done using contour integration, to give us the following expression for the partition function:
\begin{equation}
    Z_{L,M} = \frac{1}{2\pi {\rm i} }\oint \frac{{\rm d} z}{z^{M+1}}
    \tr (X_1 C^{L}) \; ,
\end{equation}
where the integration is performed along a small circle around the origin in the complex plane.
Evaluating the matrix product (which can be readily done by exploiting the fact that $C$ is lower triangular), we get:
\begin{equation}
    Z_{L,M} = \frac{1}{2\pi {\rm i} }\oint \frac{{\rm d} z}{z^{M+1}}
    \sum\limits _{l=0}^{L}\lambda _1 ^{L-l} \lambda _2 ^l \; ,
    \label{exactzlm}
\end{equation}
where $\lambda _1 = \alpha (1+z)$ and $\lambda _2 = 1+xz$ are the two eigenvalues of $C$ \eref{C}.
\subsection{Asymptotic analysis in the large $L$ limit}
By letting $w=l/L$ and approximating the sum with an integral, in the large $L$ limit, \eref{exactzlm} can be written as:
\begin{equation}
    Z_{L,M} = \frac{L}{2\pi {\rm i} }\oint \frac{{\rm d} z}{z}
   \int\limits _0 ^1 {\rm d}w \exp ( Lf(z,w) ) \; , \label{zlm_integral}
\end{equation}
where:
\begin{equation}
    f(z,w) = (1-w)\log \lambda _1 + w \log \lambda _2 - \rho \log z \; . \label{functionf}
\end{equation}
In this form, the partition function can be straightforwardly analyzed using the saddle point approach.
The partition function may be dominated either by the two-dimensional saddle point of $f$ with respect to $z$ and $w$, or by the saddle point with respect to $z$ and one of the boundary values $w=0,1$.
To check this, we first swap the order of the integrals and perform the leading order saddle point integration with respect to $z$:
\begin{equation}
    Z_{L,M} \approx \frac{L}{2 \pi {\rm i}} \int\limits _0 ^1 {\rm d}w
    \frac{1}{z_*(w)} \sqrt{\frac{2 \pi}{L |\partial _z ^2 f(z_*,w)|}}
    \exp ( Lf(z_*,w) ) \; ,
\end{equation}
where the saddle point $z_*(w)$ is defined by the equation:
\begin{equation}
    \partial _z f(z,w)|_{z=z_*} =
    \frac{1-w}{1+z_*} + \frac{xw}{1+xz_*}-\frac{\rho}{z_*}
    = 0\; . \label{zstar}
\end{equation}
This equation is quadratic but it can be checked that one of the solutions always dominates and can thus be taken to be the unique {dominant} saddle point of $f$.
It is helpful to define a reduced version of $f$:
\begin{equation}
    f_*(w) = f(z_*(w),w) \; .
\end{equation}

\begin{figure}
    \centering
    \includegraphics[scale=0.7]{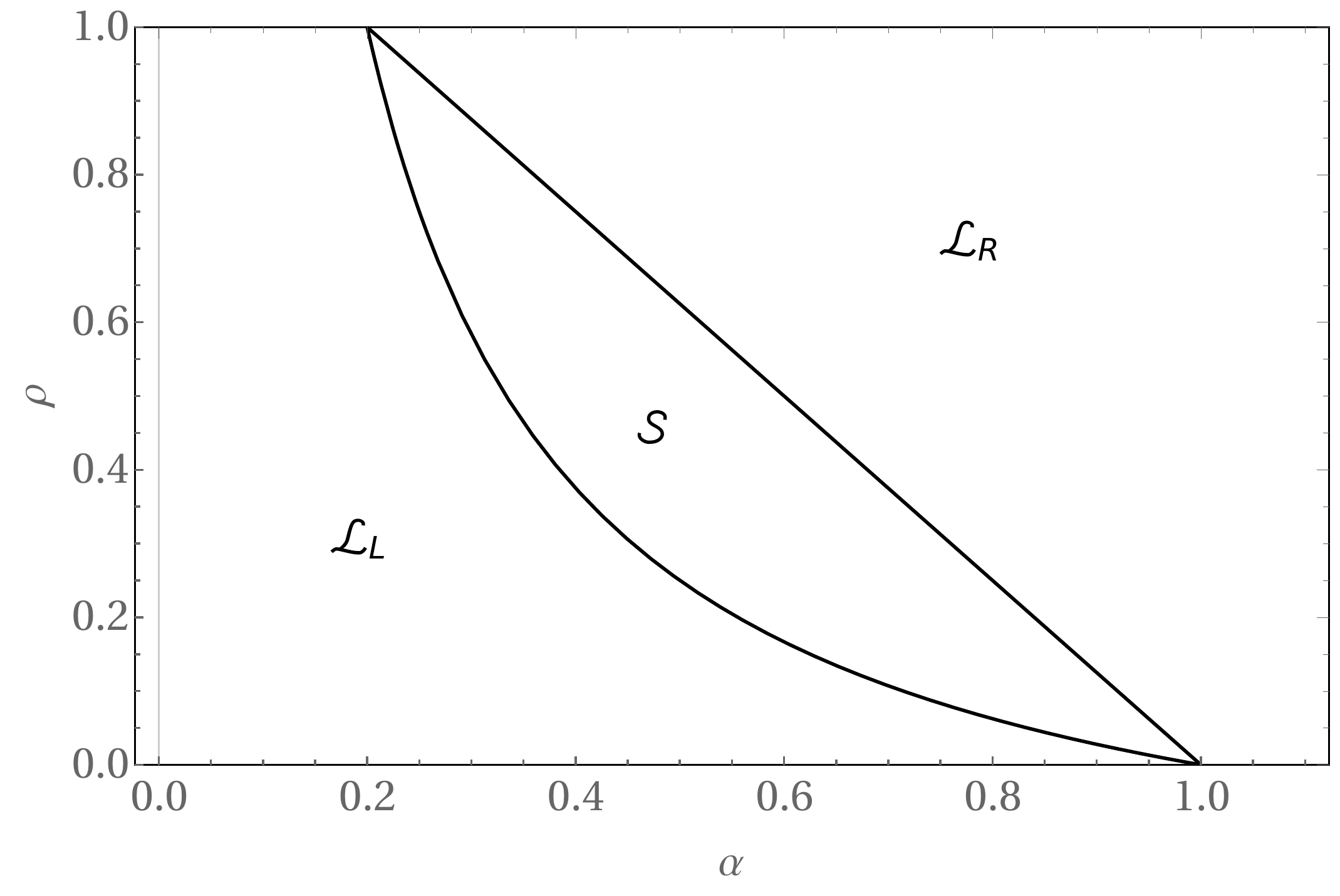}
    \caption{Phase diagram for $x=0.2$.
    The phases are labelled left/right localized ${\cal L} _{L,R}$ and shock ${\cal S}$.}
    \label{fig:phase_diagram}
\end{figure}

We now observe three different cases, which correspond to the three phases of the steady state.
On the interval $[0,1]$, $f_*(w)$ can be monotonically decreasing, increasing or have a local maximum.
In the first two cases, the partition function is dominated by the boundary values $w=0$ or $1$ respectively.
In the third case, it is dominated by the saddle point $w_* \in [0,1]$.
From \eref{zstar}, the locations of the dominant values can be found:
\numparts
\begin{eqnarray}
    w = 0 \; , \qquad z_*(0) = \frac{\rho }{1-\rho} \; ; \label{phase1}\\
    w = 1 \; , \qquad z_*(1) = \frac{\rho }{x(1-\rho )} \; ; \label{phase2}\\
    w_* = \frac{\alpha ((\alpha -1) + (1-x)\rho)}{(\alpha -1)(\alpha  -x)} \; , \qquad
    z_*(w_*) = \frac{1-\alpha}{\alpha -x} \; . \label{phase3}
\end{eqnarray}
\endnumparts
The lines of phase transition are given by the manifolds:
\begin{equation}
    \rho = \rho _1 = \frac{1-\alpha}{1-x} \; ,\qquad 
    \rho =\rho _2 = \frac{1-1/\alpha }{1-1/x} \; .
\end{equation}
These expressions agree with the mean field result for the critical densities \eref{critical_density} with $p'=\alpha p,q'=q/\alpha $.
Similarly to the mean field case, we consider fixing $p,q$ and use $\rho,\alpha $ as control parameters.
The phase diagram is shown in \fref{fig:phase_diagram}.
Defining 
\begin{equation}
    \alpha _1 = 1-\rho +\rho x \; , \qquad 
    \alpha _2 = (1-\rho +\rho /x)^{-1}\; , 
    \label{criticala} 
\end{equation}
the right and left localized phases are found in the regions $\alpha > \alpha _1$ and $\alpha < \alpha _2$ respectively, and the shock phase is found in $\alpha _2<\alpha <\alpha _1$.
We note that we always have $\alpha _1>\alpha _2$, except in the symmetric environment limit, $x\to 1$, when they are equal for all densities.
In the latter case, there is no shock phase, and in fact no phase transition.
This is shown explicitly in \sref{limits}.

\section{Density profiles and currents} \label{asymptotics}
\subsection{Density profiles}
We now derive {exact} density profiles in the reference frame of the defect in the large $L$ limit.
In the matrix product formalism, the probability of a configuration in which the $k$-th site to the right of the defect is occupied (i.e. the mean density at the $k$-th site) is:
\begin{equation}
    n _k =  [Z_{L,M}]^{-1} \frac{1}{2\pi {\rm i}} \oint \frac{{\rm d}z}{z^{M+1}}
    \tr (X_1 C^{k-1}(zX_2) C^{L-k}) \; ,
\end{equation}
where, as before, the integration is performed along a small circle around the origin to pick out the coefficient of $z^{M}$.
The trace can be evaluated as before, giving:
\begin{equation}
    n _k = 
    [Z_{L,M}]^{-1} 
    \frac{1}{2\pi {\rm i} }\oint \frac{{\rm d} z}{z^{M}} 
    \left( \alpha \sum\limits _{l=0}^{k-1}\lambda _1 ^{L-l-1} \lambda _2 ^l 
    + x \sum\limits _{l=k}^{L}\lambda _1 ^{L-l} \lambda _2 ^{l-1}
    \right) \; . \label{exactdensity}
\end{equation}
Converting to the continuous variables $w=l/L$ and $y=k/L$, we get:
\begin{equation}
    n (y) = 
    [Z_{L,M}]^{-1}
    \frac{L}{2\pi {\rm i} }\oint {\rm d}z
    \left( \frac{\alpha }{\lambda _1}\int\limits _{0}^{y} {\rm d}w 
    + \frac{x}{\lambda _2} \int\limits _{y}^{1} {\rm d}w \right) 
    \exp (Lf(z,w)) \; .
\end{equation}
This can be evaluated using the saddle point approximation, similarly to the partition function.
\subsubsection{Right localized phase}
In the right localized phase, we get the density profile:
\begin{equation}
    n _{\rm R}(y) = \rho - \frac{\rho (1-\rho )(1-x)}{\alpha _1 }
    \exp(-L \log(\alpha / \alpha _1) y) \; ,
\end{equation}
where $\alpha _1$ is as defined in \eref{criticala}.
This is a uniform density profile with a local exponentially decaying perturbation just to the right of the defect (see \fref{fig:density_plots} (c)).
The {decay} length is: 
\begin{equation}
    \xi _{\rm R} = 1/(L\log (\alpha / \alpha _1 )) \; . \label{length_r}
\end{equation}
The perturbation {from the uniform profile} is a depletion if $x<1$ and an excess if $x>1$.
\subsubsection{Left localized phase}
In the left localized phase, we get the density profile:
\begin{equation}
    n _{\rm L}(y) = \rho - \frac{\rho (1-\rho )(1-x^{-1})}{\alpha _2 }
    \exp(L \log(\alpha / \alpha _2) (1-y)) \; ,
\end{equation}
where $\alpha _2$ is as defined in \eref{criticala}.
This is similar to the right localized phase but the perturbation {from the uniform profile} is on the left side of the defect, the sign is flipped with respect to $x$ and the {decay} length is: 
\begin{equation}
    \xi _{\rm L} = 1/(L \log (\alpha /\alpha _2) ) \; , \label{length_l}
\end{equation}
(see \fref{fig:density_plots} (a)).
(The careful reader will observe that these profiles do not give exactly $\rho$ when integrated, but the error is of order $O(1/L)$, which is to be expected in a saddle point analysis.)
\subsubsection{Shock phase}
\begin{figure}
    \centering
    (a)\includegraphics[scale=0.47]{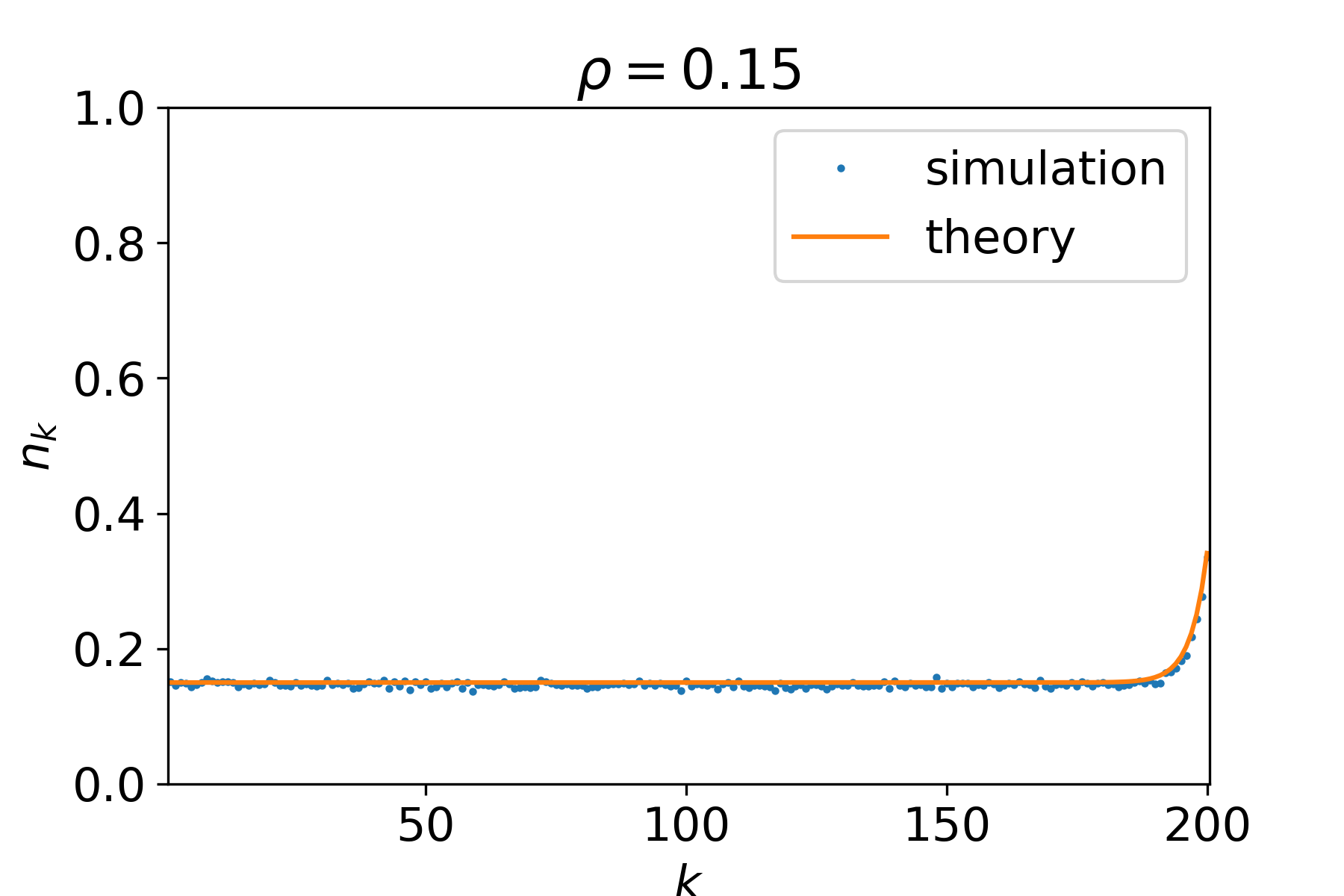}
    (b)\includegraphics[scale=0.47]{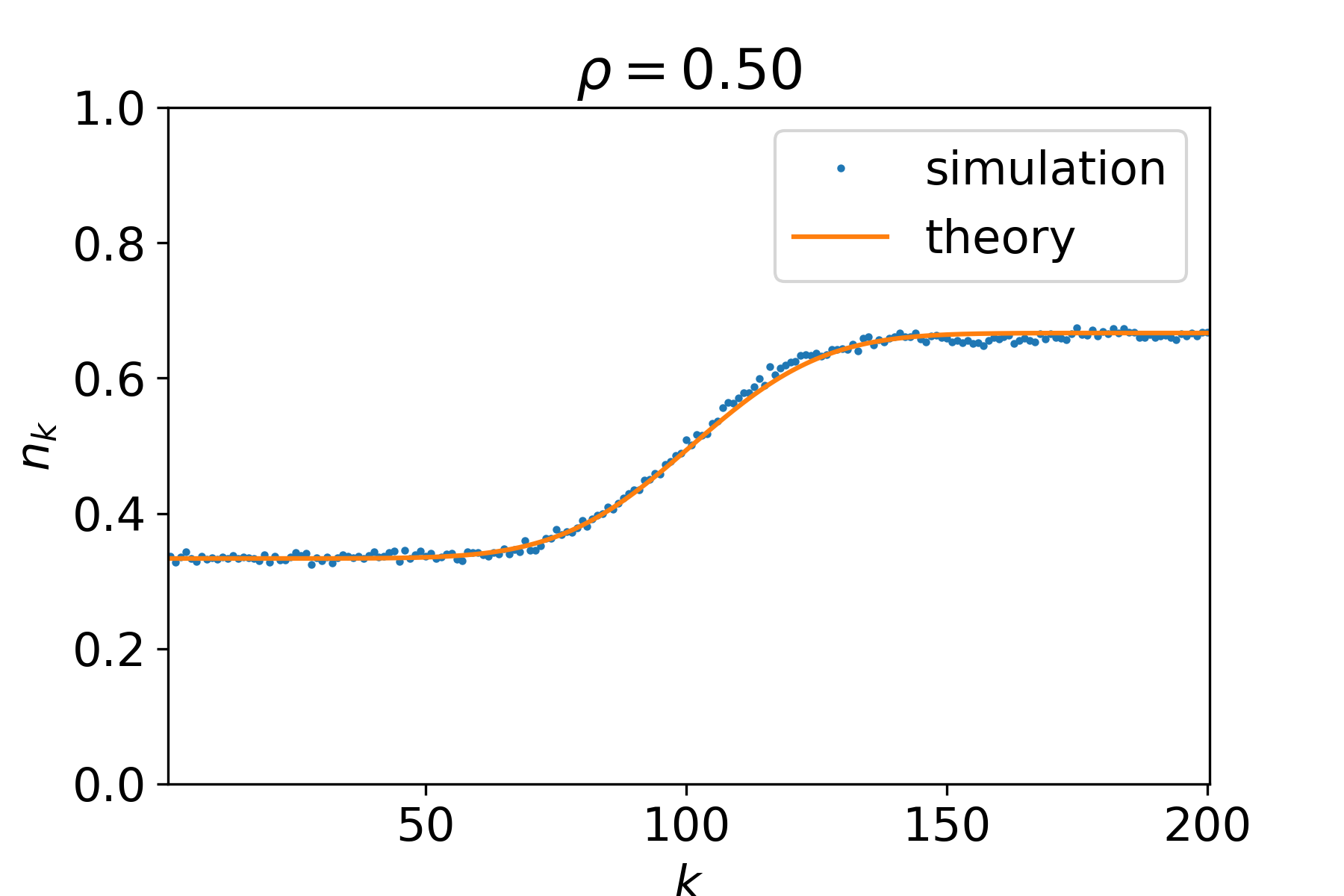}
    (c)\includegraphics[scale=0.47]{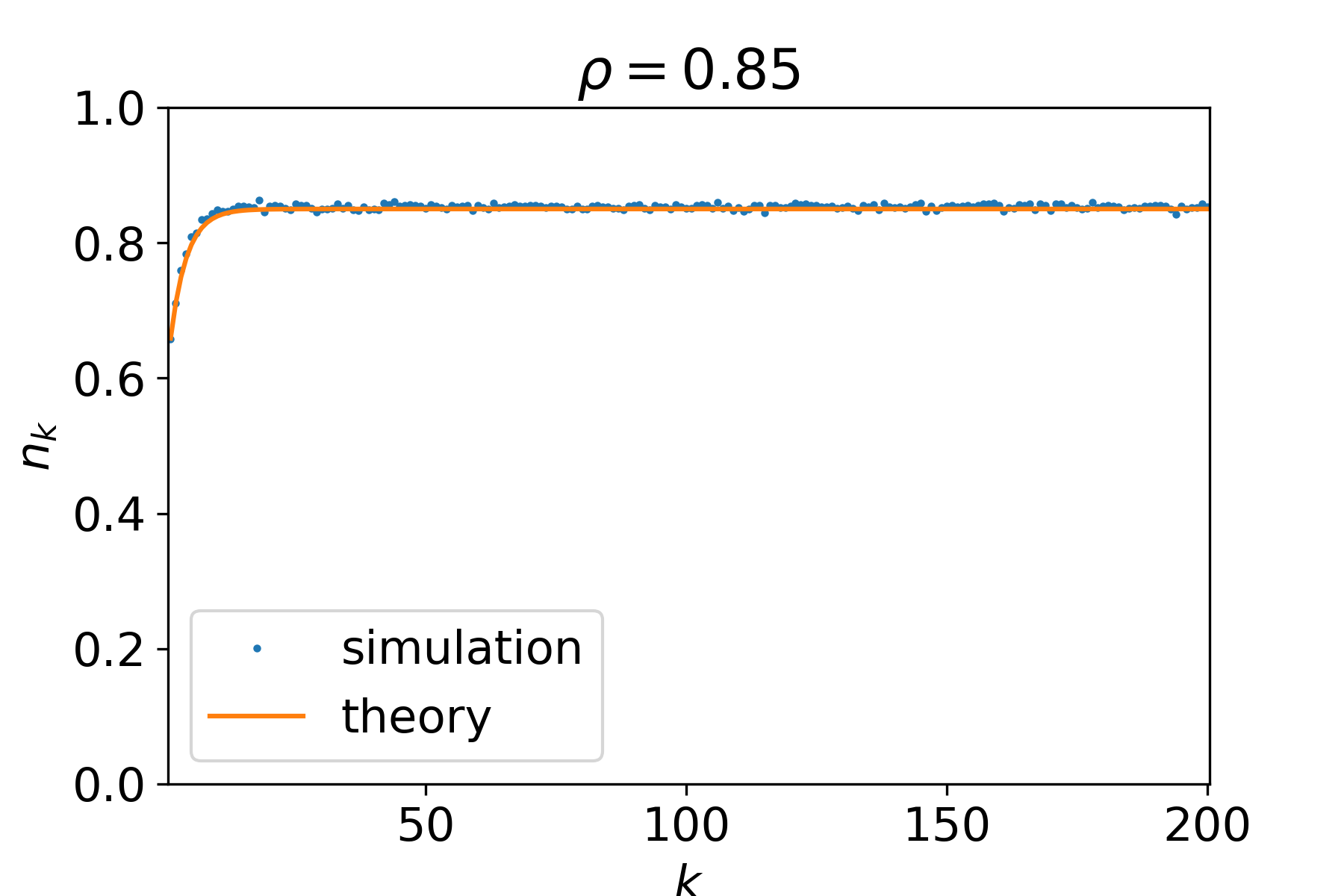}
    \caption{Density profiles in the reference frame of the defect at various densities, showing the (a) left localized, (b) shock and (c) right localized phases. Good agreement between theory and Monte Carlo simulation can be seen in all three phases. The simulations were carried out using the parameters $p=4,q=1,\alpha =0.5$ for a system size of $L=200$.}
    \label{fig:density_plots}
\end{figure}

In the shock phase, we get the density profile:
\begin{equation}
    n _{\rm S}(\zeta ) = \frac{\rho _1 + \rho _2}{2} + \frac{\rho _1-\rho _2}{2}
    {\rm erf} (\sqrt{-f_* ''(w_* )/2} \zeta ) \; ,
\end{equation}
where the scaling variable $\zeta $ is defined as $\zeta = \sqrt{L}(y-w_*)$ and we have:
\begin{equation}
    f_* ''(w_*) = 
    -\frac{(1-\alpha )^2(\alpha -x)^2}{\alpha ( (\alpha ^2 -x)(1-x)\rho + x(1-\alpha )^2 )}
    \; .
\end{equation}
This is a shock profile that interpolates between two densities, $\rho _2$ and $\rho _1$ (see \fref{fig:density_plots} (b)).
The {shock} front is centred at $y=w_*$ and has a width $\xi _{\rm S} = (-Lf_*''(w_*)/2)^{-1/2}$.
In the large $L$ limit, the width vanishes and the profile becomes a sharp shock.
Similar error function profiles with $\sqrt{L}$ scaling have been observed in the TASEP with a slow defect \cite{mallick1996shocks} and in the open boundary TASEP with a parallel sublattice updating scheme \cite{jafarpour2005exact}.

\subsection{Currents}
From the density profiles we can readily calculate the  steady-state current, $J'$, in the frame of the defect \eref{current}.
Evaluating this at leading order in the saddle point gives:
\begin{eqnarray}
    J' = 
    \left\{
    \begin{array}{ll}
        \rho (1-\rho )v -\rho v' \; , \qquad & {\rm localized \; phases} \\
        p \frac{x(1-\alpha )^2}{\alpha (1-x)} \; ,
        & {\rm shock \; phase}
    \end{array}
    \right.
    \; .
\end{eqnarray}
Then, using  \eref{galileo} we obtain the current, $J$, in the stationary frame
\begin{eqnarray}
    J = 
    \left\{
    \begin{array}{ll}
        \rho (1-\rho )v \; , & {\rm localized \; phases} \\
        \rho v' + p \frac{x(1-\alpha )^2}{\alpha (1-x)} \; , \qquad
        & {\rm shock \; phase}
    \end{array}
    \right.
    \; .
\end{eqnarray}
This agrees with the mean field result \eref{mft_current} with $p'=\alpha p,\; q'=q/\alpha $.

\begin{figure}
    \includegraphics[scale=0.55]{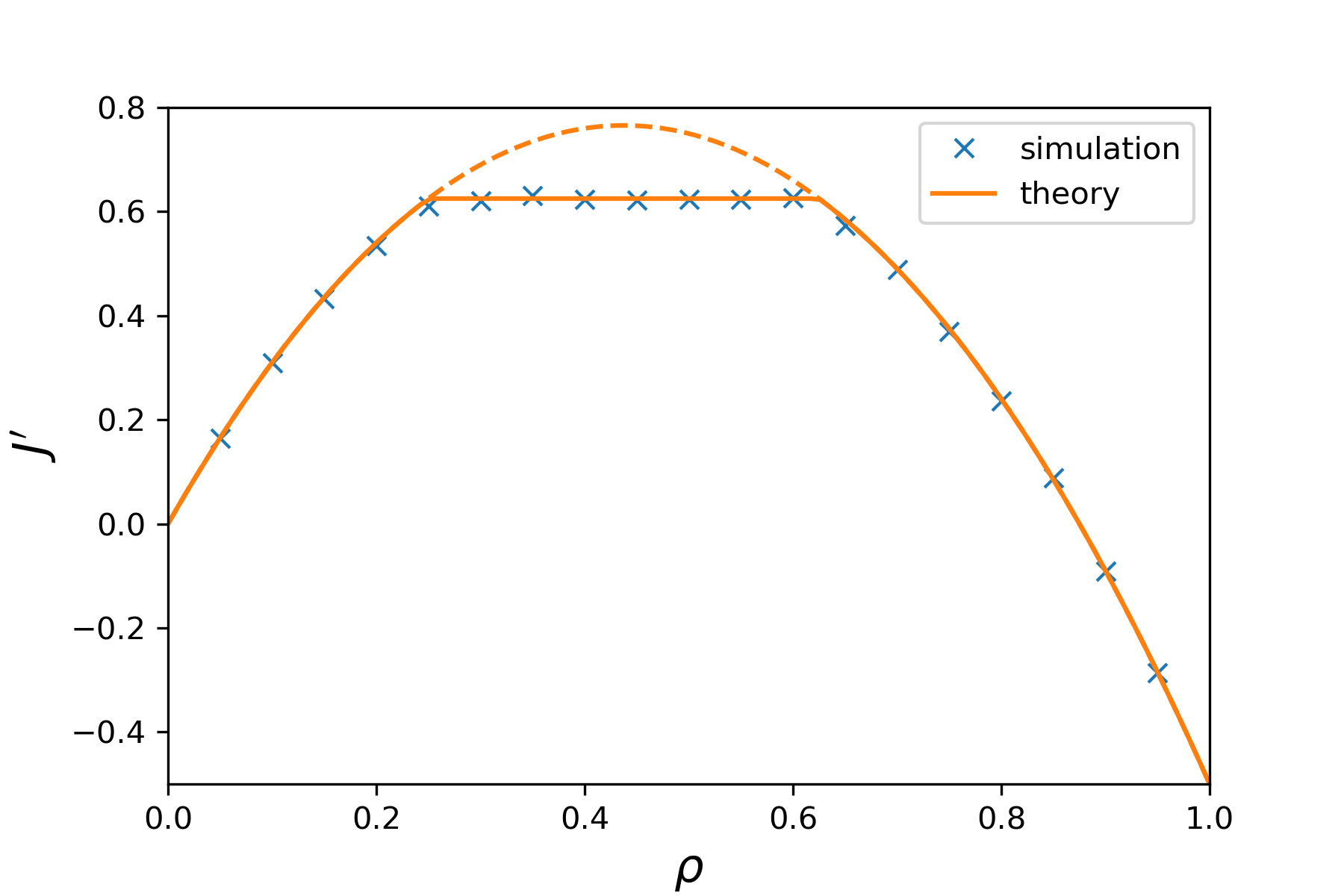}
    \includegraphics[scale=0.55]{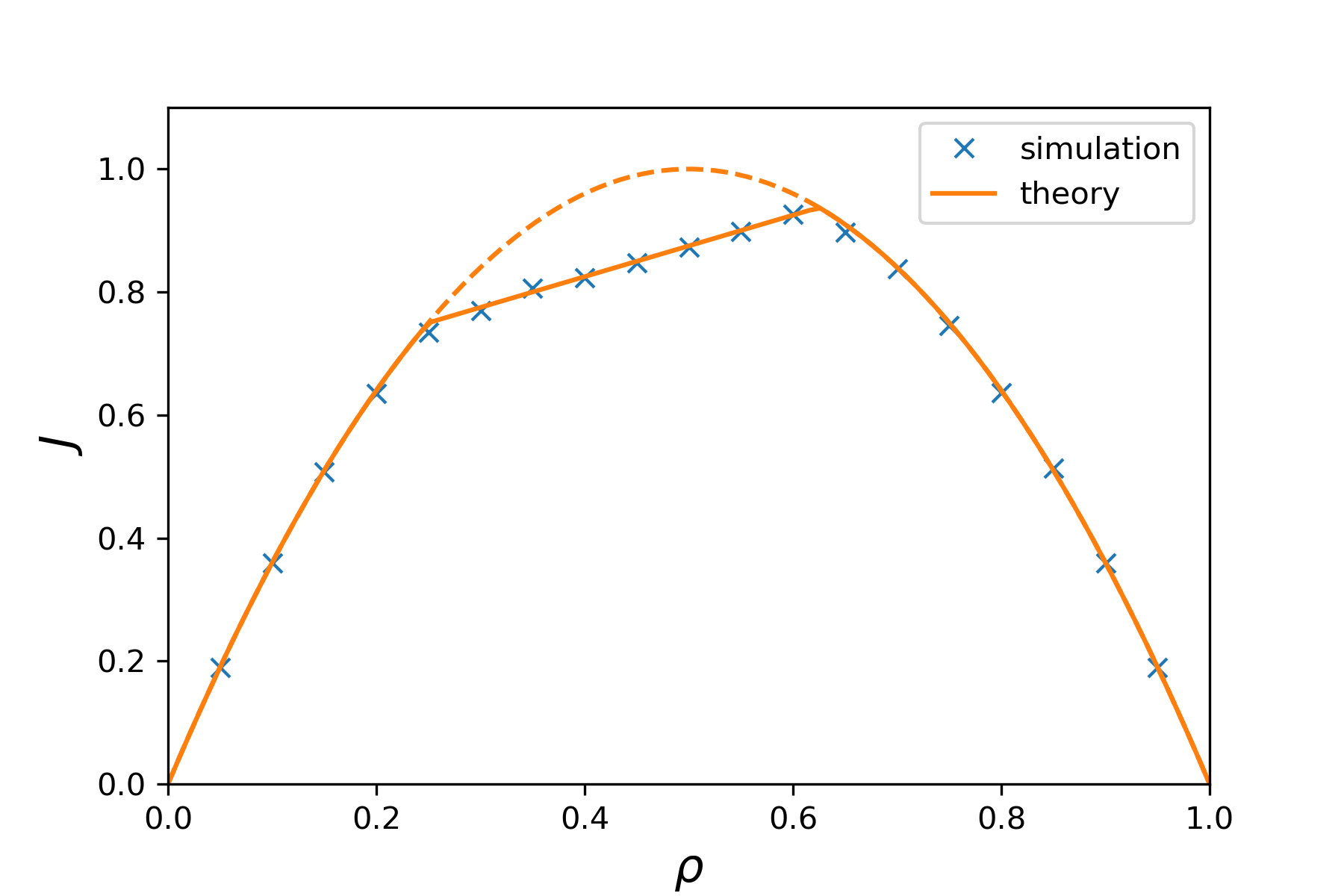}
    \caption{Current of environment particles as a function of environment density in the reference frame of the defect (left) and in the stationary frame (right). 
    The dashed curves indicate the current of a pure (defectless) PASEP.
    The currents are the same as the pure case in the localized phases and become constant (left) or linear (right) in the shock phase.
    Excellent agreement is seen between Monte Carlo simulations and theory.
    The parameters used for simulations were: $p=5,q=1,\alpha =0.5 , L=1000$.} 
    \label{fig:current}
\end{figure}
Focusing on $J'$, we note that the current is continuous but the first derivative is discontinuous at both transition
densities $\rho_1$ and $\rho_2$. Moreover,  $J'$ is constant in the shock phase  (See Fig. \ref{fig:current}).
These observations are consistent with a discontinuous phase transition between the two localized phases and the shock phase being a coexistence region between the localized phases.
It is also interesting to note that the characteristic lengths \eref{length_r}, \eref{length_l} diverge as one approaches the transition densities, since $\alpha \to \alpha _1,\; \alpha \to \alpha _2$ respectively.

\section{Exact expressions for finite-size systems} \label{exact}
The matrix product approach also allows one to derive exact expressions for the partition function and correlation functions of finite size systems in terms of combinatorial factors.
From \eref{exactzlm}, by evaluating the residue of the integrand we obtain
\begin{equation}
    Z_{L,M} = 
    \sum\limits _{l=0} ^{L} \sum\limits _{m=0} ^{l}
    {L-l \choose M-m} {l \choose m}\alpha ^{L-l} x^{m} \; . \label{zlmsum}
\end{equation}
\subsection{Density profiles and currents}
Similarly, using the residue method, we can get an exact expression for the density at site $k$ relative to the defect particle:
\begin{equation}
    n _k = Z_{L,M}^{-1}
    \left[\alpha \sum\limits _{l=0}^{k-1} \sum\limits _{m=0} ^{l}
    + x \sum\limits _{l=k-1}^{L-1} \sum\limits _{m=0} ^{l} \right]
    {L-1-l \choose M-1-m}{l \choose m}\alpha ^{L-1-l}x^{m}  \; ,
\end{equation}
and the environment current in the reference frame of the defect:
\begin{eqnarray}
    \fl
    J'=-\frac{p}{Z_{L,M}} \biggl[
    {L-1 \choose M-1}(\alpha ^{L+1}-\alpha ^{-1}x^{M+1})+ \nonumber \\
    (\alpha -1)x
    \sum\limits _{l=0}^{L-1} \sum\limits _{m=0}^{l} {L-1-l \choose M-1-m}{l \choose m}
    \alpha ^{L-1-l} x^m 
    \biggr] \; .
\end{eqnarray}
From this, the current in the stationary frame can be obtained through \eref{galileo}.
\subsection{Symmetric and totally asymmetric limits}\label{limits}
These expressions, simplify in the symmetric ($x\to 1$) and totally asymmetric ($x\to 0$) environment limits.
\begin{itemize}
    \item In the symmetric limit, $x\to 1$, the partition function becomes:
    \begin{equation}
        Z_{L,M} = {L \choose M} \frac{\alpha ^{L+1}-1}{\alpha -1} \; ,
    \end{equation}
    and the density profile is uniform: $n_k=\rho ,\;  k=1,\dots ,L$.
    It follows that the mean current of the environment is always $0$ and there are no phase transitions.
    \item In the totally asymmetric limit, $x\to 0$, the partition function becomes:
    \begin{equation}
        Z_{L,M} = \sum\limits_{l=M}^{L} {l \choose M}\alpha ^l \; ,
    \end{equation}
    and the density profile:
    \begin{equation}
        n_k = Z_{L,M}^{-1}
        \sum\limits _{l=L-k}^{L-1}{l \choose M-1}\alpha ^{l+1} \; .
    \end{equation}
    Evaluating the asymptotics, we get a localized phase for $\alpha >1-\rho $ and a shock phase with bulk densities $\rho _1=1-\alpha ,\rho _2 = 0$ for $\alpha < 1-\rho $.
    The second class particle current is $p\rho (1-\rho )$ in the localized phase and $\alpha p \rho $ in the shock phase.
\end{itemize}

\section{Conclusion} \label{conclusion}
We have considered a PASEP with a generalized first class defect.
This can be seen as a generalization of defect problems that have been previously considered in literature, like the defect in a TASEP \cite{derrida1993exact,mallick1996shocks} and a driven defect in a SSEP \cite{burlatsky1992directed,burlatsky1996motion,miron2020phase,ayyer2020simple,lobaskin2020driven}.
The phase diagram was obtained through mean field theory and for a special case through the matrix product ansatz.
This brings the range of solved models one step closer to the most general driven defect problem (with independent environment, defect and overtaking rates).

The phase diagram (\fref{fig:mft_phase}, \fref{fig:phase_diagram}) comprises a low-density localized phase ($\mathcal{L}_L$) and a high-density localized phase ($\mathcal{L}_R$), separated by a shock phase ($\mathcal{S}$).
The shock phase can be interpreted as a region of phase coexistence between the high- and low-density phases, in analogy to a liquid-gas system.
In this light, one expects the transition to be discontinuous, as there is a density jump (the shock) between the two coexisting phases.
This is borne out by the current in the reference frame of the defect ($J'$).
$J'$ has to be equal in the two coexisting phases for the shock to have a fixed average position in the moving frame.
Our analysis shows that $J'$ is continuous at the transitions and its first derivative is discontinuous, which is consistent with a discontinuous phase transition.
Interestingly, there are also diverging length scales at the transitions, associated with the size of the localized density perturbations due to the defect.

A key difference between our results and the symmetric case is that in the symmetric case, no shock phase is observed \cite{miron2020phase,lobaskin2020driven,ayyer2020simple}.
There, the effects of the defect are localized or extend through the whole system, but the extended phase does not have two separate bulk densities.

It is interesting to note that the matrices presented here do not form a closed algebra that allows matrix product reductions to be carried out (see \ref{reductions} for an argument for this).
Therefore they fall outside of the previously known classification of simple matrix product states \cite{isaev2001diffusion}.
It would be of interest to investigate whether this result can be generalized to a wider class of solvable models.

It is also worth noting the matrices \eref{matrices} cannot be used to describe the steady state of a system with more than one defect.
It would be of interest to investigate whether a matrix product solution can be found for that case.
Such a generalization would also be needed in order to treat the open boundary version of this model.
It has been suggested that there exists a connection between Yang-Baxter integrable models and models with a simple matrix product steady state, with a mapping between the two being derived for some cases \cite{crampe2014integrable,sasamoto1997stationary}.
In a future publication we will present strong evidence that the case with more than one defect is not Yang-Baxter integrable but the single defect case can be integrated using a nested coordinate Bethe ansatz \cite{tobepublished}.

\section*{Acknowledgements}
IL acknowledges studentship funding from EPSRC under Grant No. EP/R513209/1.
The work of KM has been supported by the project RETENU ANR-20-CE40-0005-01 of the French National Research Agency (ANR).
KM thanks Tomohiro Sasamoto for stimulating discussions on integrable exclusion processes, which informed this work.
MRE would like to thank David Mukamel for helpful discussions.

\appendix
\section{Proof of validity of matrix product ansatz} \label{proof}
We now provide a proof that the matrices given in \eref{matrices} produce weights that form the steady state.
Previous proofs have exploited relations similar to (\ref{algebra10}-\ref{algebra20}), which allow for the matrix products to be reduced \cite{blythe2007nonequilibrium}.
As noted earlier, in the case at hand, matrix product reductions are not possible, so the proof does not follow automatically.
However, due to the simplicity of the representation, it can be checked directly.

We consider the weight of some configuration $\{ \tau _1,\dots ,\tau _{L+1} \}$ given by the corresponding matrix product.
For ease of notation, we define site swapping operators:
\begin{equation}
    \fl
    \hat{h}_{i,i+1} f(\{ \dots \tau _i, \tau _{i+1} \dots \} ) =
    w_{i+1,i} f(\{ \dots \tau _{i+1}, \tau _{i} \dots \} ) \nonumber
    - w_{i,i+1} f(\{ \dots \tau _i, \tau _{i+1} \dots \} ) \; ,
\end{equation}
where $w_{i,i+1}$ indicates the rate of the process $\tau _i \tau_{i+1} \to \tau _{i+1} \tau_{i}$.
The two terms correspond to the fluxes in and out of the configuration due to processes which swap $\tau _i,\tau _{i+1}$.
Then the condition for the steady state is given by:
\begin{equation}
    \sum\limits _{i=1}^{L+1} \hat{h}_{i,i+1} f(\{\tau _1,\dots ,\tau _{L+1} \}) = 0 \; . \label{master}
\end{equation}
Focusing on the effect of two consecutive site swapping operators, we get:
\begin{eqnarray}
    \fl
    (\hat{h}_{i-1,i} + \hat{h}_{i,i+1} ) 
    & f(\{ \dots \tau _{i-1},\tau_{i},\tau_{i+1} \dots \}) = 
    \tr (\dots [w_{i,i-1} X_{\tau _{i}} X_{\tau_{i-1}} 
    - w_{i-1,i} X_{\tau _{i-1}} X_{\tau_{i}} ]X_{\tau_{i+1}} \dots ) \nonumber \\
    & + \tr (\dots X_{\tau_{i-1}}[w_{i+1,i} X_{\tau _{i+1}} X_{\tau_{i}} 
    - w_{i,i+1} X_{\tau _{i}} X_{\tau_{i+1}} ] \dots ) \; .
\end{eqnarray}
We now use the substitutions (\ref{algebra10}-\ref{algebra20}) on the terms in the square brackets and isolate the terms not containing the matrix $X_{\tau_i}$ in the resulting expressions.
If $\tau _i \neq 1$, they are: 
\begin{equation}
    \tr (\dots [x_{\tau _i} X_{\tau _{i-1}}] X_{\tau _{i+1}} \dots )
    - \tr (\dots X_{\tau _{i-1}} [x_{\tau _i} X_{\tau _{i+1}}] \dots ) \; ,
\end{equation}
which cancel since $x_{\tau _i}$ is scalar.
This type of cancellation is the only possibility one needs to consider for matrices which allow product reductions.
Here we also have $\tau _i = 1$, which gives:
\begin{equation}
    \tr ( \dots Y_{\tau_{i-1}} X_{\tau _{i+1} } \dots )
    - \tr ( \dots X_{\tau _{i-1} } Y_{\tau _{i+1}} \dots ) \; .
\end{equation}
It can be checked that for all possibilities $\tau _{i-1},\tau_{i+1}=0,2$, in the matrix given by the expression $(Y_{\tau_{i-1}} X_{\tau _{i+1} } - X_{\tau _{i-1} } Y_{\tau _{i+1}})$, only the lower left entry is non-zero.
As the remaining matrices in that product will all be $X_0$ or $X_2$, it follows that the product as a whole is traceless and these terms cancel as well.
Repeating this argument for all $i$, we get that the weights given by the matrix product satisfy \eref{master}.

\section{Non-existence of matrix product reduction relations} \label{reductions}
We present a more rigorous argument that the matrix product solution described in this paper does not form a closed algebra that allows matrix product reductions to be carried out.
For a general two species problem with transition rates $\omega _{\tau \tau'}\equiv W(\tau \tau'\to \tau' \tau)$, one expects matrices $X_0,X_2,X_2$ that satisfy relations of the form:
\begin{equation}
    \omega _{\tau \tau '}X_{\tau}X_{\tau '} - \omega _{\tau ' \tau}X_{\tau '} X_{\tau}  =
    x_{\tau '}X_\tau -x _{\tau } X_{\tau '} \; , \label{commutator}
\end{equation}
for $\tau ,\tau ' = 0,1,2$ and where $x _{\tau},x_{\tau '}$ are some scalars to be determined \cite{blythe2007nonequilibrium}.
Note that this relation reduces the expression from a product of two matrices to single matrices.
One typically exploits relations of this type to evaluate matrix products by reducing them to a simple expression, which eliminates the need for an explicit representation.
In order for this ansatz to be a consistent description of the steady state, it has to satisfy a diamond lemma, which can be expressed as the following conditions for the model parameters \cite{isaev2001diffusion}:
\begin{eqnarray}
    x_1 \omega _{02}(\omega _{12}-\omega_{21}-\omega _{10}+\omega _{01}) = 0 \; , 
    \label{automatic} \\
    x_2 \omega _{01}(\omega _{12}-\omega_{21}+\omega _{20}-\omega _{02}) = 0 \; , \\
    x_0 \omega _{21}(\omega _{20}-\omega_{02}-\omega _{10}+\omega _{01}) = 0 \; , \\
    x_2 x_2(\omega _{12}-\omega_{21}+\omega _{20}-\omega _{10}) = 0 \; , \\
    x_1 x_0 (\omega _{21} - \omega _{02}) =0\; , \\
    x_2 x_0 (\omega _{10} - \omega _{12} -\omega _{20} + \omega _{02}) = 0 \; .
\end{eqnarray}
In the model considered in this paper, we have:
\begin{equation}
    \fl 
    \omega _{10}=\omega_{12} = \alpha p \; ; \qquad
    \omega _{01} = \omega _{21} = q/\alpha  \; ; \qquad 
    \omega _{20} = p \; ; \qquad  \omega _{02} = q \; .
\end{equation}
Putting these parameters in, we get that only the LHS of \eref{automatic} vanishes automatically.
To satisfy the remaining conditions, we must set $x_0=x_2=0$.
This would imply that $pX_2 X_0-qX_0 X_2=0$.
Using this relation, one could take the weight of any configuration of the system and move all the $X_2$'s to the left, ending up with an expression of the form $(p/q)^N \tr (X_1 X_2^M X_0^{L-M})$, for some integer $N$.
Thus weights of different configurations can only differ by powers of $p/q$.
This can easily seen to be insufficient by diagonalizing the master equation exactly for small system sizes.
Thus there exist no choice of matrices $X_0,X_1,X_2$ and scalars $x_0,x_1,x_2$ that satisfy \eref{commutator} for this system.

\section*{References}
\bibliographystyle{iopart-num.bst}
\bibliography{references}

\providecommand{\newblock}{}
\begin{thebibliography}{10}
\expandafter\ifx\csname url\endcsname\relax
  \def\url#1{{\tt #1}}\fi
\expandafter\ifx\csname urlprefix\endcsname\relax\def\urlprefix{URL }\fi
\providecommand{\eprint}[2][]{\url{#2}}

\bibitem{wolf1996traffic}
Wolf D~E, Schreckenberg M and Bachem A 1996 {\em Traffic and granular flow\/}
  (World Scientific)

\bibitem{cividini2017driven}
Cividini J, Mukamel D and Posch H~A 2017 {\em Physical Review E\/} {\bf 95}
  012110

\bibitem{SNCM2018}
Szavits-Nossan J, Ciandrini L and Romano M~C 2018 {\em Phys. Rev. Lett.\/} {\bf
  120}(12) 128101

\bibitem{scott2019power}
Scott S and Szavits-Nossan J 2019 {\em Physical biology\/} {\bf 17} 015004

\bibitem{derrida1993exact}
Derrida B, Janowsky S~A, Lebowitz J~L and Speer E~R 1993 {\em Journal of
  statistical physics\/} {\bf 73} 813--842

\bibitem{janowsky1992finite}
Janowsky S~A and Lebowitz J~L 1992 {\em Physical Review A\/} {\bf 45} 618

\bibitem{derrida1997shock}
Derrida B, Lebowitz J~L and Speer E~R 1997 {\em Journal of statistical
  physics\/} {\bf 89} 135--167

\bibitem{mallick1996shocks}
Mallick K 1996 {\em Journal of Physics A: Mathematical and General\/} {\bf 29}
  5375

\bibitem{jafarpour2005exact}
Jafarpour F~H, Ghafari F~E and Masharian S~R 2005 {\em Journal of Physics A:
  Mathematical and General\/} {\bf 38} 4579

\bibitem{tabatabaei2006shocks}
Tabatabaei F and Sch{\"u}tz G~M 2006 {\em Physical Review E\/} {\bf 74} 051108

\bibitem{evans1996}
Evans M~R 1996 {\em Europhysics Letters\/} {\bf 36} 13--18

\bibitem{blythe2007nonequilibrium}
Blythe R~A and Evans M~R 2007 {\em Journal of Physics A: Mathematical and
  Theoretical\/} {\bf 40} R333

\bibitem{CMZ2011}
Chou T, Mallick K and Zia R~K~P 2011 {\em Reports on Progress in Physics\/}
  {\bf 74} 116601

\bibitem{burlatsky1992directed}
Burlatsky S~F, Oshanin G~S, Mogutov A~V and Moreau M 1992 {\em Physics Letters
  A\/} {\bf 166} 230--234

\bibitem{miron2020phase}
Miron A, Mukamel D and Posch H~A 2020 {\em Journal of Statistical Mechanics:
  Theory and Experiment\/} {\bf 2020} 063216

\bibitem{derrida1992exact}
Derrida B, Domany E and Mukamel D 1992 {\em Journal of statistical physics\/}
  {\bf 69} 667--687

\bibitem{dehp1993}
Derrida B, Evans M~R, Hakim V and Pasquier V 1993 {\em Journal of Physics A:
  Mathematical and General\/} {\bf 26} 1493

\bibitem{schutz1993phase}
Sch{\"u}tz G and Domany E 1993 {\em Journal of statistical physics\/} {\bf 72}
  277--296

\bibitem{janowsky1994exact}
Janowsky S~A and Lebowitz J~L 1994 {\em Journal of Statistical Physics\/} {\bf
  77} 35--51

\bibitem{basu2016passage}
Basu R, Sidoravicius V and Sly A 2016 Last passage percolation with a defect
  line and the solution of the slow bond problem (\textit{Preprint}
  \eprint{1408.3464})

\bibitem{burlatsky1996motion}
Burlatsky S~F, Oshanin G, Moreau M and Reinhardt W~P 1996 {\em Physical Review
  E\/} {\bf 54} 3165

\bibitem{lobaskin2020driven}
Lobaskin I and Evans M~R 2020 {\em Journal of Statistical Mechanics: Theory and
  Experiment\/} {\bf 2020} 053202

\bibitem{ayyer2020simple}
Ayyer A to be published {\em Annales de l'Institut Henri Poincar{\' e} D\/}

\bibitem{cividini2018driven}
Cividini J, Mukamel D and Posch H 2018 {\em Journal of Physics A: Mathematical
  and Theoretical\/} {\bf 51} 085001

\bibitem{sahoo2014dynamic}
Sahoo M, Dong J and Klumpp S 2014 {\em Journal of Physics A: Mathematical and
  Theoretical\/} {\bf 48} 015007

\bibitem{sahoo2016asymmetric}
Sahoo M and Klumpp S 2016 {\em Journal of Physics A: Mathematical and
  Theoretical\/} {\bf 49} 315001

\bibitem{szavits2020current}
Szavits-Nossan J and Waclaw B 2020 {\em Physical Review E\/} {\bf 102} 042117

\bibitem{sasamoto2000one}
Sasamoto T 2000 {\em Physical Review E\/} {\bf 61} 4980

\bibitem{wood2020combinatorial}
Wood A~J, Blythe R~A and Evans M~R 2020 {\em Journal of Physics A: Mathematical
  and Theoretical\/} {\bf 53} 123001

\bibitem{tobepublished}
Lobaskin I, Evans M~R and Mallick K to be published

\bibitem{isaev2001diffusion}
Isaev A~P, Pyatov P~N and Rittenberg V 2001 {\em Journal of Physics A:
  Mathematical and General\/} {\bf 34} 5815

\bibitem{crampe2014integrable}
Crampe N, Ragoucy E and Vanicat M 2014 {\em Journal of Statistical Mechanics:
  Theory and Experiment\/} {\bf 2014} P11032

\bibitem{sasamoto1997stationary}
Sasamoto T and Wadati M 1997 {\em Journal of the Physical Society of Japan\/}
  {\bf 66} 2618--2627

\end{thebibliography}
\end{document}